\documentclass{article}
\begin{document}

    \title{\bf Covariant hamiltonian formalism \\ for field theory:
    Hamilton-Jacobi equation on the space $\cal G$}
    
\author{\bf Carlo Rovelli\\[1mm] 
{\em Centre de Physique Th\'eorique
de Luminy, CNRS,}\\ {\em Case 907, 
 F-13288 Marseille, EU}} 
\date{\today}
    
    \maketitle

    \begin{abstract} 
	\noindent Hamiltonian mechanics of field theory can be
	formulated in a generally covariant and background independent
	manner over a \emph{finite} dimensional extended configuration
	space.  The physical symplectic structure of the theory can
	then be defined over a space $\cal G$ of three-dimensional
	surfaces without boundary, in the extended configuration
	space.  These surfaces provide a preferred
	over-coordinatization of phase space.  I consider the
	covariant form of the Hamilton-Jacobi equation on $\cal G$,
	and a canonical function $S$ on $\cal G$ which is a preferred
	solution of the Hamilton-Jacobi equation.  The application of
	this formalism to general relativity is equivalent to the ADM
	formalism, but fully covariant.  In the quantum domain, it
	yields directly the Ashtekar-Wheeler-DeWitt equation. 
	Finally, I apply this formalism to discuss the partial
	observables of a covariant field theory and the role of the
	spin networks --basic objects in quantum gravity-- in the
	classical theory.
    \end{abstract}
    
   \vskip1cm 
   
\section{Introduction}

Hamiltonian mechanics is a clean and general formalism for describing
a physical system, its states and its observables, and provides a road
towards quantum theory.  In its traditional formulation, however, the
hamiltonian formalism is badly non covariant.  This is a source of
problems already for finite dimensional systems.  For instance, the
notions of state and observable are not very clean in the hamiltonian
mechanics of the systems where evolution is given in parametric form,
especially if the evolution cannot be deparametrized (as in certain
cosmological models).  But the problem is far more serious in field
theory.  The conventional field theoretical hamiltonian formalism
breaks manifest Lorentz invariance badly.  More importantly, in a
generally covariant context the conventional hamiltonian formalism is
cumbersome and its physical interpretation is far from being
transparent.

In my opinion, a proper understanding of the generally covariant
structure of mechanics is necessary in order to make progress in
quantum gravity.  The old notions of observable, state, evolution,
hamiltonian, and so on, and the tools that are conventionally employed
to relate quantum field theory with experience --$S$ matrix, $n$-point
functions and so on, cease to make sense in a genuinely general
covariant and background independent context.  Therefore wee need to
understand the the generally covariant version of these notions.  In
particular, if we want to understand quantum field theory in a truly
background independent context, we must find a proper definition of
quantum field theoretical transition amplitudes, in a form with a
clear operational interpretation that makes sense also when there is
no background spacetime.

A covariant formulation of the hamiltonian mechanics of finite
dimensional systems is possible.  Several versions of this formulation
can be found in the literature, with different degrees of
developments.  Perhaps the first to promote this point of view was
Lagrange himself, who first understood that the proper covariant
definition of phase space is as the space of the physical motions
\cite{Lagrange}, or the space of the solutions of the equations of
motion (modulo gauges).  Notable contributions, among many others, are
Arnold's identification of the presymplectic space with coordinates
$(t,q^i,p_{i})$ (time, lagrangian varia bles and their momenta) as the
natural home for mechanics \cite{Arnold}, and the beautiful and well
developed, but little known, formalism of Souriau \cite{Souriau}. 
Here, I use the covariant version of hamiltonian mechanics described
in reference \cite{I}, which builds on previous results.  This
formalism is based on the physical notion of ``partial observable"
\cite{partial}.  The partial observables of a non-relativistic finite
dimensional mechanical system are the quantities $(t,q^i)$, treated on
the same footing.  In particular, the formalism treats the time
variable $t$ on the same footing as the lagrangian variables $q^i$. 
The space of the partial observables is the extended configuration
space $\cal C$ and the hamiltonian formalism is built over this space. 
The phase space $\Gamma$ is identified with a space of one-dimensional
curves in $\cal C$.  The elements of $\cal C$ and $\Gamma$ provide the
proper relativistic generalization of the notions of observable and
state, consistent with the modification of the notions of space and
time introduced by general relativity \cite{I}.  The formalism can be
manifestly Lorentz covariant and deals completely naturally with
reparametrization invariant systems such as the cosmological models.

The extension of these ideas to field theory require the
identification of the partial observables of field theory.  These are
finite in number.  They include the coordinates of the spacetime $M$
and the coordinates of the target space $T$ on which the fields take
value.  Thus the extended configuration space of a field theory is the
finite dimensional space ${\cal C}=M\times T$.  A hamiltonian
formalism for field theory built on the finite dimensional space
${\cal C}=M\times T$ has been developed by many authors in several
variants, developing classical works by Cartan, Weyl \cite{W} and
DeDonder \cite{DD} on the calculus of variations.  See for instance
\cite{hf} and especially the beautiful work \cite{M} and the extended
references therein.  Here, I refer to the version of hamiltonian
mechanics for field theory described in \cite{II}, where the accent is
on general covariance and relation to observability.  The (infinite
dimensional) phase space $\Gamma$ is identified as a space of
four-dimensional surfaces in $\cal C$.  The physical symplectic form
of $\Gamma$, which determines the Poisson brackets, was not given in
\cite{II}.  Here, I consider a definition of the symplectic form on
$\Gamma$, in this context.  (On a covariant definition of the
symplectic structure on the space of the solutions of the field
equations, see also \cite{AW}, whose relation with this work will be
briefly discussed below.)

The key of the construction is to introduce the space $\cal G$.  The
space $\cal G$ is the space of the boundary --``initial and
final"--lagrangian data (no momenta).  In the finite dimensional case,
a typical element $\gamma$ of $\cal G$ is a pair of points in $\cal
C$.  Generically, two points in $\cal C$ --say $(t_{0},q^i_{0})$ and
$(t,q^i)$-- identify a motion.  In the field theoretical case, $\cal
G$ is a space of three-dimensional surfaces $\gamma$, without
boundary, in $\cal C$.  Again, $\gamma$ is a set of Lagrangian data
sufficient to identify a motion.  Indeed $\gamma$ defines a closed
hypersurface in spacetime and the value of the fields over it.  I show
below that there is a canonical two-form $\omega_{\cal G}$ on $\cal
G$, and the physical phase space and its physical symplectic form,
namely its Poisson brackets, follow immediately from the pair $({\cal
G},\omega_{\cal G})$.

I then study a covariant version of the Hamilton-Jacobi theory,
defined on $\cal G$, and I observe that there exist a preferred,
canonical, solution $S[\gamma]$ of the Hamilton-Jacobi equation on
$\cal G$.  The Hamilton-Jacobi formalism is a window open towards
quantum theory.  Schr\"odinger introduced the Schr\"odinger equation
by interpreting the Hamilton-Jacobi equation as the optical
approximation of a wave equation \cite{Sh2}.  This means searching for
an equation for a wave function $\psi$, solved to lowest order in
$\hbar$ by $\psi=Ae^{i/\hbar S}$, if $S$ solves the Hamilton-Jacobi
equation.  On the basis of this idea, Schr\"odinger found his
celebrated equation by simply replacing each partial derivative of $S$
in the Hamilton-Jacobi function with (-$i\hbar$ times) a partial
derivative operator \cite{Sh1}.  This same procedure can be used in
the covariant formulation of mechanics.  The covariant Hamilton-Jacobi
equation yields then directly the quantum dynamical equation of the
theory.  This is the Schr\"odinger equation in the case of a
non-covariant system, or the appropriate ``Wheeler-DeWitt" equation
for covariant systems.  (In fact, the Wheeler-DeWitt equation as well
was first found replacing partial derivatives with partial derivative
operators in the Hamilton-Jacobi equation of general relativity
\cite{brice}.)  For a parametrized system, this procedure shortcuts
Dirac's recipe for the quantization of first class constraints, which
is cumbersome and has a very cloudy interpretation when applied to
such systems.  Furthermore, if $S[\gamma]$ is the canonical solution
of the Hamilton-Jacobi equation mentioned above, then
$\psi[\gamma]=A[\gamma]e^{i/\hbar S[\gamma]}$ is the propagator of the
Schr\"odinger equation, which was identified in \cite{reisenberger} as
the quantity providing a direct operational interpretation to a
generally covariant quantum system.

In the field theoretical context, several authors have developed a
covariant Hamilton-Jacobi formalism based the Weyl-DeDonder approach. 
In this formalism the Hamilton-Jacobi equation is a partial
differential equation for a multiplet of Hamilton-Jacobi functionals
defined on the extended configuration space.  (For a review, see
\cite{Ka}).  Here I do not utilize this formalism, because I do not
see how it could lead to the quantum theory (on this, see \cite{k}). 
Instead, I discuss a Hamilton-Jacobi function over $\cal G$, which
provides a general covariant setting for a {\em functional\/}
Hamilton-Jacobi equation.

I then apply this formalism to general relativity (following also
\cite{gr4} and \cite{II}.)  I use self-dual variables, which much
simplify the equations \cite{selfdual,lee}.  The extended
configuration space is identified with the finite-dimensional space
$\tilde{\cal C}=M\times {\cal C}$, where $M$ is the four-dimensional
manifold of the spacetime coordinates and ${\cal C}=R^4\times sl(2,C)$
where $sl(2,C)$ is the Lorentz algebra.  The formalism is simple and
straightforward.  It shortcuts the intricacies of the conventional
hamiltonian formalism of general relativity, and further simplify the
one described in \cite{II}.  The Hamilton-Jacobi equation yields
immediately to the Ashtekar form of the Wheeler-DeWitt equation.  (See
extended references in \cite{report}.  On early use of Hamilton-Jacobi
theory in general relativity, see \cite{HJGR}.)  In the classical as
well as in the quantum theory, the $M$ (spacetime) component of
$\tilde{\cal C}$ ends up playing only an auxiliary role and it
disappears from observables and states.  This reflects the
diffeomorphism invariance of the theory.

Finally, I discuss the physical interpretation of the formalism.  This
is of interest for the interpretation of quantum gravity and, more in
general, any background independent quantum field theory.  The
physical predictions of the theory are given in terms of correlations
between partial observables.  These can be obtained directly from
$S[\gamma]$.  In the context of general relativity, this leads to the
introductions of spin networks in the classical context, opening a
bridge towards the spinnetworks used in the quantum theory (spin
networks were introduced in quantum gravity in \cite{sn} and then
developed in \cite{sna}.  See \cite{loops} and extended references
therein.)

The paper is structured as follows.  The main notions are introduced in Section
\ref{finite} in the finite dimensional context.  For concreteness, I exemplify
all structures introduced by computing them explicitly in the simple case of a
free particle.  The field theoretical formalism is developed in Section
\ref{field}.  As an example, I describe a self-interacting scalar field in
Minkowski space.  General relativity is treated in Section \ref{GR}.  In this
paper, I am not concerned with global issues: I deal only with aspects of the
theory which are local in $\cal C$.

\section{Finite dimensional systems}\label{finite}

The exercise that we perform in this Section, following and extending the
results of \cite{I}, is to carefully reformulate classical hamiltonian mechanics
and its interpretation in a form that does not require any of the
non-relativistic notions that loose meaning in a generally covariant context. 
These notions are for instance: instantaneous state of the system, evolution in
time, evolution of the observables in time, and so on.  The alternative notion
that we use (correlations, motions \ldots) are introduced in detail in 
\cite{I}. 
The motivation of this exercise is to introduce the ideas that will then be used
for field theory in Section \ref{field} and for general relativity in Section
\ref{GR}.

\subsection{Relativistic mechanics}

Consider a system with $n$ degrees of freedom governed by a Hamiltonian function
$H_{0}(t,q^i,p_{i})$, where $q^i$ with $i=1,\ldots,n$, are coordinates on the
configuration space, $p_{i}$ the corresponding momenta and $t$ the time
variable.  Let $\cal C$ be the ($n$+1)-dimensional extended configuration space,
that is, the product of the configuration space with the real line, with
coordinates $q^a=(t,q^i)$, with $a=0,\ldots,n$.  From now on we work on this
extended configuration space and we treat $t$ on the same footing as the
configuration space variables.  In this form mechanics is general enough to
describe fully relativistic systems.

Let $\Omega=T^*{\cal C}$ be the cotangent space to $\cal C$, and
$p_a=(p_{0},p_i)\equiv(\pi,p_i)$ the momenta conjugate to $q^a$.  Being a
cotangent space, $\Omega$ carries the canonical one-form
$\theta_{\Omega}=p_{a}dq^a$.  Dynamics is defined on $\Omega$ by the
relativistic hamiltonian (or hamiltonian constraint)
\begin{equation}
	   H(q^a,p_{a})=\pi +H_{0}(q^a,p_{i})=0.
	   \label{H}
\end{equation}
This equation defines a surface $\Sigma$ in $\Omega$.  It is convenient to
coordinatize $\Sigma$ with the coordinates $(q^a,p_i)=(t,q^i,p_i)$.  We denote
$\theta$ the restriction of $\theta_{\Omega}$ to $\Sigma$ and $\omega=d\theta$. 
An orbit of $\omega$ is a curve without boundaries
$m:\tau\to(q^a(\tau),p_i(\tau))$ in $\Sigma$ whose tangent $X=(\dot
q^a\partial_{a}+\dot p_{i}\partial^i)$ satisfies
\begin{equation}
\omega(X)=0.
\label{omegaX}
\end{equation}
Here the dot indicates the derivative with respect to $\tau$ while
$\partial_{a}$ and $\partial^i$ form the basis of tangent vectors in $\Sigma$
associated with the coordinates $(q^a,p_i)$.  The last equation is equivalent to
the Hamilton equations.  The projection $\tilde m$ of $m$ to $\cal C$, that is
the curve $\tilde m:\tau \to q^a(\tau) = (t(\tau),q^i(\tau))$, is a solution of 
the
equations of motion, given in parametric form.

Let $\Gamma$ be the space of the orbits.  There is a natural projection
\begin{equation}
\Pi :\Sigma\to\Gamma \label{pi}
\end{equation}
that sends a point to the orbits to which it belongs.  There is also a unique
two-form $\omega_{\Gamma}$ on $\Gamma$ such that its pull back to $\Sigma$ is 
$\omega$:
\begin{equation}
\Pi_*\omega_{\Gamma}=\omega.
\label{omegaomegaG}
\end{equation}
The symplectic space $(\Gamma,\omega_{\Gamma})$ is the physical phase space with
its physical symplectic structure. 

This formalism, as well as its interpretation, can be immediately generalized to
the case in which the coordinates $q^a$ of $\cal C$ do {\em not\/} split into
$t$ and $q^i$ and the relativistic hamiltonian does {\em not\/} have the
particular form (\ref{H}).  Therefore it remains valid for generally covariant
or reparametrization invariant systems \cite{I}.

\subsubsection*{Example: free particle}

Consider a free particle in one dimension.  Then $n=1$, the extended
configuration space has coordinates $t$ (the time) and $x$ (the position of the
particle).  Denote $\pi$ and $p$ the corresponding momenta.  The constraint
(\ref{H}) that defines the free motion is
\begin{equation}
	   H=\pi +\frac{1}{2m}p^2=0.
\label{Hfree}
\end{equation}
The restriction of the canonical form $\theta_{\omega}=\pi dt+pdx$ to the
surface $\Sigma$ defined by (\ref{Hfree}) is
\begin{equation}
	   \theta=-\frac{p^2}{2m} dt + pdx
\end{equation}
where we have taken the coordinates $(t,x,p)$ for $\Sigma$.  The two form
$\omega$ on $\Sigma$ is therefore
\begin{equation}
 \omega=d\theta=-\frac{p}{m} dp\wedge dt + dp \wedge dx= dp\wedge\left(dx-
 \frac{p}{m} dt\right).
\label{omegafp}
 \end{equation}
A curve $(t(\tau),x(\tau),p(\tau))$ on $\Sigma$ has tangent $X=\dot
t\partial_{t}+\dot x\partial_{x}+\dot p\partial_{p}$ and inserting this and
(\ref{omegafp}) in the
main equation (\ref{omegaX}) we obtain
\begin{eqnarray}
\omega(X)=-\dot t \frac{p}{m} dp +\dot x dp - \dot p \left(dx- \frac{p}{m}
dt\right)=0.
\end{eqnarray}
Equating to zero each component, we have
\begin{eqnarray}
 \dot x = \frac{p}{m} \dot t  , \ \ \ \ \ 
\dot p = 0 .
\end{eqnarray}
With the solution $x(\tau)=\frac{p}{m} t(\tau)+Q$ and $p(\tau)=P$ where $Q$ and
$P$ are constants and $t(\tau)$ arbitrary.  The projection of this orbit on
$\cal C$ gives the motions $x=\frac{P}{m} t+Q$.

The space of the orbits is thus parametrized by the two integration constants
$Q$ and $P$.  The point $(t,x,p)$ in $\Sigma$, belongs to the orbit
$Q=x-\frac{p}{m} t$ and $P=p$.  Thus the projection (\ref{pi}) is given by $\Pi
(t,x,p)=(Q(t,x,p),P(t,x,p))=(x-\frac{p}{m} t,p)$.  Then
$\omega_{\Gamma}=dP\wedge dQ$, because
\begin{eqnarray}
\Pi_*\omega_{\Gamma} &=& dP(t,x,p)\wedge dQ(t,x,p) \\ \nonumber 
 &=& dp \wedge d\left(x-\frac{p}{m} t\right)
\ =\ dp \wedge \left(dx-\frac{p}{m} dt\right)
\ =\ \omega,
\end{eqnarray}
as required by the definition (\ref{omegaomegaG}).  See \cite{I} for examples of
relativistic systems.

\subsection{The space $\cal G$}

We now introduce a space which is important for what follows.  Let $\cal G$ be
defined as
\begin{equation}
{\cal G} = {\cal C} \times {\cal C}.
\end{equation}
That is, an element $\gamma$ of $\cal G$ is an ordered pair of elements of the
extended configuration space $\cal C$:
\begin{equation}
\gamma = (q^a,q_0^a)= (t,q^i,t_{0},q^i_0).
\end{equation}
We think at $\gamma$ as initial and final conditions for a physical motion: the
motion begins at $q^i_{0}$ at time $t_{0}$ and ends at $q^i$ at time $t$. 
Generically, given $\gamma= (q^a,q_0^a)$ there is a unique solution of the
equations of motion that goes from $q_{0}^a$ to $q^a$.\footnote{More precisely,
we define $\cal G$ as the set of pairs for which this solution exists.  If there
is more that one solution, we choose the one with minimal action.  See below.}
That is, there is a curve $(q^a(\tau),p_{i}(\tau))$, with boundaries, in
$\Sigma$, with $\tau\in[0,1]$ such that 
\begin{eqnarray}
q^a(0) = q_{0}^a,  \ \ \ \ \ 
q^a(1) = q^a,  \ \ \ \ \ 
\omega(X)=0.
\end{eqnarray}
We denote $m_{\gamma}$ this curve, $\tilde m_{\gamma}$ its projection to $\cal
C$ (namely $q^a(\tau)$).  Thus $\gamma$ is the boundary of $\tilde m_{\gamma}$,
which we write as $\gamma=\partial \tilde m_{\gamma}$.  We denote $s$ and
$s_{0}$ the initial and final points of $m_{\gamma}$ in $\Sigma$.  Notice that
$s=(q^a,p_{i})$ and $s_{0}=(q_{0}^a,p_{0}{}_{i})$, where in general $p_{i}$ and
$p_{0}{}_{i}$ depend both on $q^a$ as well as on $q^a_{0}$.

There is a natural map $i:{\cal G}\to\Gamma$ which sends each pair to the orbit
that the pair defines.  Thus we can define a two-form $\omega_{\cal G}$ on $\cal
G$ as $\omega_{\cal G}=i^*\omega_{\Gamma}$.  In other words,
$\gamma=(q^a,q^a_{0})=(t,q^i,t_{0},q^i_{0})$ can be taken as a natural
(over-)coordinatization of the phase space.  Instead of coordinatizing a motion
with initial positions and momenta, we coordinatize it with initial and final
positions.  In these coordinates, the symplectic form is given by $\omega_{\cal
G}$.

For what follows, it is important to notice that there is an equivalent, 
alternative definition of $\omega_{\cal G}$, which can be obtained without going
first through $\Gamma$.  Indeed, let $\delta\gamma=(\delta q^a, \delta 
q^a_{0})$ be a vector (an infinitesimal displacement) at $\gamma$. Then the 
following is true:
\begin{eqnarray}
\omega_{\cal G}(\gamma)(\delta_{1}\gamma,\delta_{2}\gamma) &=& \omega_{\cal
G}(q^a,q^a_{0})((\delta_{1}q^a,\delta_{1}q^a_{0}), 
(\delta_{2}q^a,\delta_{2}q^a_{0}))
\nonumber \\
&=& \omega(s)(\delta_{1}s,\delta_{2}s) -
\omega(s_{0})(\delta_{1}s_{0},\delta_{2}s_{0}).
\end{eqnarray}
Notice that $\delta_{1}s$, the variation of $s$, is determined by $\delta_{1}q$
as well as by $\delta_{1}q_{0}$, and so on.  This equation expresses
$\omega_{\cal G}$ directly in terms of $\omega$.  As we shall see, this equation
admits an immediate generalization in the field theoretical framework, where
$\omega$ will be a five-form, but $\omega_{\cal G}$ is a two-form.

Now fix a pair $\gamma= (q^a,q^a_0)$ and consider a small variation of only one
of its elements.  Say
\begin{equation}
\delta\gamma = (\delta q^a,0).
\end{equation}
This defines a vector $\delta\gamma$ at $\gamma$ on $\cal G$, which can be
pushed forward to $\Gamma$.  If the variation is along the direction of the
motion, then the push forward vanishes, that is $i_{*}\delta\gamma=0$, because
$\gamma$ and $\gamma+\delta\gamma$ define the same motion.  It follows that if
the variation is along the direction of the motion, $\omega_{\cal
G}(\delta\gamma)=0$.  Thus clearly the equation
\begin{equation}
\omega_{\cal G}(X)=0.
\end{equation}
gives again the solutions of the equations of motion.

Thus, the pair $({\cal G}, \omega_{\cal G})$ contains all the relevant
information on the system.  The null directions of $\omega_{\cal G}$ define the
physical motions, and if we divide ${\cal G}$ by these null directions, the
factor space is the physical phase space, equipped with the physical symplectic
structure.

\subsubsection*{Example: free particle}

The space $\cal G$ has coordinates $\gamma=(t,x,t_{0},x_{0})$.  Given this point
in $\cal G$, there is clearly one motion that goes from $(t_{0},x_{0})$ to
$(t,x)$, which is
\begin{eqnarray}
t(\tau)&=& t_{0}+(t-t_{0})\tau, \\
x(\tau)&=& x_{0}+(x-x_{0})\tau.
\end{eqnarray}
Along this motion,
\begin{eqnarray}
p&=& m\frac{x-x_{0}}{t-t_{0}}, \\
\pi&=& -\frac{(x-x_{0})^2}{2m(t-t_{0})^2}.
\end{eqnarray}
The map $i:{\cal G}\to\Gamma$ is thus given by
\begin{eqnarray}
P&=& p = m\frac{x-x_{0}}{t-t_{0}}, \label{P} \\
Q&=& x-\frac{p}{m} t = x-\frac{x-x_{0}}{t-t_{0}}t, \label{Q}
\end{eqnarray}
and therefore the two-form $\omega_{\cal G}$ is
\begin{eqnarray}
\omega_{\cal G}&=&i^*\omega_{\Gamma}
= dP(t,x,t_{0},x_{0})\wedge dQ(t,x,t_{0},x_{0}) \nonumber \\
&=& m\ d\frac{x-x_{0}}{t-t_{0}}\wedge d\left(x-\frac{x-x_{0}}{t-t_{0}}t\right)
\nonumber \\
&=& \frac{m}{t-t_0}\left(dx-\frac{x-x_{0}}{t-t_{0}}dt \right) \wedge
\left(dx_{0}-\frac{x-x_{0}}{t-t_{0}}dt_{0}\right).
\end{eqnarray}
It is immediate to see that a variation $\delta\gamma=(\delta t, \delta x,0,0)$ 
(at
constant $(x_{0},t_{0})$) such that $\omega_{\cal G}(\delta\gamma)=0$ must
satisfy
\begin{eqnarray}
\delta x = \frac{x-x_{0}}{t-t_{0}}\ \delta t .
\end{eqnarray}
This is precisely a variation of $x$ and $t$ along the physical motion
(determined by $(x_{0},t_{0})$).  Therefore $\omega_{\cal G}(\delta\gamma)=0$
gives again the equations of motion.  The two null directions of $\omega_{\cal
G}$ are thus given by the two vector fields
\begin{eqnarray}
X&=&\frac{x-x_{0}}{t-t_{0}}\partial_{x}+\partial_{t},\\
X_{0}&=&\frac{x-x_{0}}{t-t_{0}}\partial_{x_{0}}+\partial_{t_{0}},
\end{eqnarray}
which are in involution (their Lie bracket vanishes), and therefore define a
foliation of $\cal G$ with two-dimensional surfaces.  These surfaces are
parametrized by $P$ and $Q$, given in (\ref{P},\ref{Q}), and in fact
\begin{eqnarray}
X(P)=X(Q)=X_{0}(P)=X_{0}(Q)=0.
\end{eqnarray}
In fact, we have simply recovered in this way the physical phase space: the
space of these surfaces is the phase space $\Gamma$ and the restriction of
$\omega_{\cal G}$ to it is the physical symplectic form $\omega_{\Gamma}$.

\subsection{The function $S$ on the space $\cal G$: a preferred solution of the
Hamilton-Jacobi equation}

Let us now introduce an important function on $\cal G$.  We define
$S(\gamma)=S(q^a,q^a_{0})$ by 
\begin{equation}
S(\gamma)=\int_{m_{\gamma}} \theta.
\label{Sgamma}
\end{equation}
It is easy to see that this is the value of the action along the path
$\tilde m_{\gamma}$.  In fact
\begin{eqnarray}
S(\gamma)&=&\int_{m_{\gamma}} \theta =\int_{m_{\gamma}} p_{a}dq^a \\ 
\nonumber 
&=&\int_0^1 p_{a}(\tau)\dot q^a(\tau)d\tau =\int_0^1 \left(\pi (\tau)\dot
t(\tau)+p_{i}(\tau)\dot q^i(\tau)\right)d\tau \nonumber \\
&=&\int_0^1 \left(-H_{0}(\tau)\dot t(\tau)+p_{i}(\tau)\dot q^i(\tau)\right)d\tau
=\int_{t_0}^t \left(-H_{0}(\tau)+p_{i}(\tau)\frac{dq^i(t)}{dt}\right)dt 
\nonumber
\\
&=&\int_{t_0}^t L\left(q^i,\frac{dq^i(t)}{dt}\right) dt,
\end{eqnarray}
where $L$ is the Lagrangian.  We have from the definition (\ref{Sgamma}) of 
$S$, 
\begin{eqnarray}
\frac{\partial S(q^a,q^a_{0})}{\partial q^a}=p_{a}(q^a,q^a_{0})
\label{key}
\end{eqnarray}
where $p_{a}$ is the value of the momenta in $s$ (which depends on $q^a$
\emph{and} $q^a_{0}$).  The derivation of this equation is less obvious than
what it looks at first sight: see appendix A. 

It follows from  (\ref{key}) that $S$ satisfies the (covariant)
Hamilton-Jacobi equation \cite{I}
\begin{eqnarray}
H\left(q^a,\frac{\partial S(q^a,q^a_{0})}{\partial q^a}\right)=0.
\end{eqnarray}
More precisely, $S$ satisfies the Hamilton-Jacobi equation in both sets of
variables, namely it satisfies also
\begin{eqnarray}
H\left(q_{0}^a,-\frac{\partial S(q^a,q^a_{0})}{\partial q_{0}^a}\right)=0,
\end{eqnarray}
where the minus sign comes from the fact that the second set of variable is in
the lower integration boundary in (\ref{Sgamma}).  $S(\gamma)$, defined in
(\ref{Sgamma}), is thus a preferred solution of the Hamilton-Jacobi solution.

In view of the generalization to field theory, it is convenient to extend the
definition of $\cal G$ and $S(\gamma)$ as follow.  Define
\begin{equation}
S(q^a_{f},q^a,q^a_{i})=\int_{m_{q^a_{f},q^a}}
\theta+\int_{m_{q^a,q^a_{i}}} \theta.
\label{Sgamma3}
\end{equation} 
This can be seen as an extension of the definition (\ref{Sgamma}) in the
following sense.  We can view the path $m=m_{q^a_{f},q^a}\cup m_{q^a,q^a_{i}}$
as a path in $\cal C$ bounded by the three points $(q^a_{f},q^a,q^a_{i})$.  This
form of $S$ will be convenient for extracting physical information from it, and
for the extension to field theory.

\subsubsection*{Example: free particle}

The function $S(\gamma)$ is easily computed.
\begin{eqnarray}
S(t,x,t_{0},x_{0}) &=& \int_{0}^1 (\pi\dot t+p\dot x) 
= \pi\int_{t_O}^t dt+ p\int_{x_0}^x dx 
\nonumber \\
&=& -\frac{m(x-x_{0})^2}{2(t-t_{0})} + m\frac{(x-x_{0})^2}{t-t_{0}} 
\nonumber \\
&=& \frac{m(x-x_{0})^2}{2(t-t_{0})}.
\label{actionfp}
\end{eqnarray}
We now verify that it satisfies the Hamilton-Jacobi equation of the 
relativistic 
hamiltonian (\ref{Hfree}), which is 
\begin{eqnarray}
H\left(q^a,\frac{\partial S}{\partial q^a} \right)= \frac{\partial S}{\partial
t}+ \frac{1}{2m}\left(\frac{\partial S}{\partial x}\right)^2=0.
\label{HJp}
\end{eqnarray}
Easily
\begin{eqnarray}
\frac{\partial S}{\partial t}= -\frac{m(x-x_{0})^2}{2(t-t_{0})^2}, 
&\hspace{1cm}&
\frac{\partial S}{\partial x}= \frac{m(x-x_{0})}{(t-t_{0})}.
\end{eqnarray}
So that the Hamilton-Jacobi equation (\ref{HJp}) is satisfied.  

Notice that $S(x,t,x_{0},t_{0})$ is strictly related to the quantum theory.  The
Shr\"odinger equation can be obtained from the relativistic Hamiltonian-Jacobi
equation as 
\begin{eqnarray}
 H\left(q^a,-i\hbar\frac{\partial}{\partial q^a}\right) \psi(q^a)=0.
\end{eqnarray}
for a wave function $\psi(q^a)$ on the extended phase space.  In fact, this
gives immediately
\begin{eqnarray}
\left(-i\hbar\frac{\partial}{\partial
t}-\frac{\hbar^2}{2m}\frac{\partial^2}{\partial x^2} \right)\psi(x,t)=0.
\end{eqnarray}
which is the conventional time dependent Shr\"odinger equation.  Its propagator
$W(x,t,x_{0},t_{0})$, which satisfies the equation itself
\begin{eqnarray}
\left(-i\hbar\frac{\partial}{\partial
t}-\frac{\hbar^2}{2m}\frac{\partial^2}{\partial x^2} \right)
W(x,t,x_{0},t_{0})=0 
\end{eqnarray}
is 
\begin{eqnarray}
W(x,t,x_{0},t_{0})=\frac{1}{\sqrt{t-t_{0}}}\
e^{\frac{i}{\hbar}S(x,t,x_{0},t_{0})}.
\end{eqnarray}
Therefore $S$ is the phase of the propagator.

\subsection{Physical predictions}

Finally, let us discuss the relation between the formalism described and the
physical interpretation of the theory.  If the function $S(\gamma)$ on $\cal G$
is known explicitly, the general solution of the equation of motion is
\begin{eqnarray}
f^a(q^a,q_{0}^a,p^0_{a})= \frac{\partial S(q^a,q^a_{0})}{\partial
q^a_{0}}+p^0_{a}=0.
\label{HJsol}
\end{eqnarray}
We view (\ref{HJsol}) as an equation for $q^a$; the quantities $p^0_{a}$ and
$q_{0}^a$ are constants determining one solution.  The solution defines a curve
in $\cal C$, namely a motion.  Physically, once we have determined a motion by
means of $p^0_{a}$ and $q_{0}^a$, equation (\ref{HJsol}) determines whether or
not the correlation $q^a$ can be observed.  (See \cite{I} for more details.)  In
general, there is a redundancy in the system (\ref{HJsol}): one of the equations
($a=0$ for non relativistic systems) is a consequence of the others.

For instance, in the case of a free particle, inserting (\ref{actionfp}) into 
(\ref{HJsol}) gives, for $q^a=x$ 
\begin{eqnarray}
\frac{\partial S(t,x,t_{0},x_{0})}{\partial x_{0}}+p^{0}=0.
\end{eqnarray}
Inserting the explicit form of $S$ given in (\ref{actionfp}), we obtain
\begin{eqnarray}
 x-x_{0}=\frac{p^0}{m}(t-t_{0}).
\end{eqnarray}
which is the correct relation that relates $x$ and $t$, at constant values of
$x_{0},t_{0},p^{0}$.  The other equation, obtained for $q^a=t$, does not give
anything new.

There is another way of using the solution of the Hamilton-Jacobi to obtain
physical predictions, which is of interest in view of it generalization to
quantum field theory.  Fix two points $q_{i}^a$ and $q_{f}^a$ in $\cal C$.  We
can ask if a third point $q^a$ can lay on the same motion as $q_{i}^a$ and
$q_{f}^a$.  This means asking wether or not we could observe the
correlation $q^a$, given that the correlations $q_{i}^a$ and $q_{f}^a$ are
observed.  A moment of reflection will convince the reader that the answer to
this question is positive if and only if
\begin{eqnarray}
\frac{\partial S(q^a_{f},q^a_{i})}{\partial q^a_{f}}
=\frac{\partial S(q^a_{f},q^a)}{\partial q^a_{f}}, \\ 
\frac{\partial S(q^a_{f},q^a_{i})}{\partial q^a_{i}}
=\frac{\partial S(q^a,q^a_{1})}{\partial q^a_{i}}.
\label{thirdpoint}
\end{eqnarray}
Using the definition (\ref{Sgamma3}), this becomes
\begin{eqnarray} 
\frac{\partial S(q^a_{f},q^a,q^a_{i})}{\partial q^a_{\tau}}
=\frac{\partial S(q^a_{f},q^a_{i})}{\partial q^a_{\tau}},
\label{thirdpoint2}
\end{eqnarray}
with $\tau=i,f$. 

Alternatively, we may notice that a first order variation of $q^a$ does not
change $S(q^a_{2},q^a,q^a_{1})$, because the two derivatives of the first and
second term in (\ref{Sgamma3}) cancel out if $m$ is a motion (and therefore the
two momenta at $q^a$ are equal).  Thus the condition can be reformulated as
\begin{eqnarray} 
\frac{\partial S(q^a_{f},q^a,q^a_{i})}{\partial q^a}=0. 
\label{thirdpoint4}
\end{eqnarray}
This equation is easily recognized as a corollary of the principle that the
action is an extremum on the motions.

Consider now the same question --whether or not we can observe the correlation
$q^a$-- in the quantum theory.  The quantum theory does not provide a
deterministic yes/no answer to a physical question; it only provides the
probability (amplitude) for a positive answer.  In the classical theory, the two
correlations $q_{i}^a$ and $q_{f}^a$ determine a state.  In the quantum theory,
a (generalized) state of a finite dimensional system is determined by a single
point in $\cal C$, because the detection of the correlation $q_{i}^a$ is
incompatible with the detection of the correlation $q_{f}^a$, in the sense of
Heisenberg: the detection of the second erases the information on the first.  If
the state is the generalized state determined by the correlation $q^a_{1}$, then
the probability density amplitude for detecting the correlation $q^a$ is
determined by the propagator $W(q^a,q^a_{i})$, defined on $\cal G$.  See
\cite{reisenberger} for details.  Since $S(q^a,q^a_{i})$ is the phase of the
propagator, equations (\ref{thirdpoint}) follow then from the standard optical
approximation for the behavior of the wave packets.  In this sense the classical
interpretation of the formalism can be recovered from the quantum one.

\section{Field theory}\label{field}

\subsection{Field theoretical relativistic mechanics}

Consider a field theory on Minkowski space $M$.  Let $x^\mu$, where
$\mu=0,1,2,3$, be Minkowski coordinates and call $\phi^A(x^\mu)$ the field,
where $A=1, \ldots, N$.  The field is a function $\phi:M\to T$, where
$T=I\!\!R^N$ is the target space, namely the space in which the field takes
values.  The extended configuration space of this theory is the finite
dimensional space ${\cal C}= M\times T$, with coordinates $q^a=(x^\mu,\phi^A)$. 
In fact, the coordinates of this space correspond to the (4+$N$) partial
observables whose relations are described by the theory \cite{partial,II}.  A
solution of the equations of motion defines a four-dimensional surface $\tilde
m$ in $\cal C$.  If we coordinatize this surface using the coordinates $x^\mu$,
then this surface is given by $[x^\mu,\phi^A(x^\mu)]$, where $\phi^A(x^\mu)$ is
a solution of the field equations.  If, alternatively, we use an arbitrary
parametrization with parameters $\tau^\rho, \rho=0,1,2,3$, then
the surface is given by $[x^\mu(\tau^\rho),\phi^A(\tau^\rho)]$, and 
$\phi^A(x^\mu)$ is determined by $\phi^A(x^\mu(\tau^\rho))=\phi^A(\tau^\rho)$.

In the case of a finite numbe of degrees of freedom (and no gauges), motions are
given by one-dimensional curves.  At each point of the curve, there is one
tangent vector, and momenta coordinatize the one-forms.  In field theory,
motions are four-dimensional surfaces, and have four independent tangents at
each point.  Accordingly, momenta coordinatize the four-forms.  Let
$\Omega=\Lambda^4T^*{\cal C}$, be the bundle of the four-forms
$p_{abcd}dq^a\wedge dq^b \wedge dq^c \wedge dq^d$ over $\cal C$.  A point in
$\Omega$ is thus a pair $(q^a,p_{abcd})$.  The space $\Omega$ carries the
canonical four-form $\theta_{\Omega}=p_{abcd}\, dq^a\wedge dq^b \wedge dq^c
\wedge dq^d$.  It is convenient to use the notation $p_{\mu\nu\rho\sigma} =
\pi\epsilon_{\mu\nu\rho\sigma}$ and $p_{A\nu\rho\sigma} =
p_{A}^\mu\epsilon_{\mu\nu\rho\sigma}$.

The hamiltonian theory can be defined on $\Omega$ by the relativistic
hamiltonian system 
\begin{eqnarray}
p_{ABCD}=p_{ABC\mu}=p_{AB\mu\nu}&=&0, \label{constr1} \\ 
H= \pi + H_{0}(x^\mu,\phi^A,p_{A}^\mu)&=&0. \label{constr2}
\end{eqnarray}
where $H_{0}$ is DeDonder's covariant Hamiltonian \cite{DD} (see below for an
example).  This system defines a surface $\Sigma$ in $\Omega$.  It is convenient
to take coordinates $(x^\mu,\phi^A,p_{A}^\mu)$ on $\Sigma$.  As before, we
denote $\theta$ the restriction of $\theta_{\Omega}$ to $\Sigma$ and
$\omega=d\theta$.  On the surface defined by (\ref{constr1}), $\theta_{\Omega}$
becomes the canonical four-form
\begin{equation}
\theta= \pi\ d^4x + p_{A}^\mu\ d\phi^{A}\wedge d^3x_{\mu}, 
\end{equation}
where we have introduced the notation $d^4x=dx^0\wedge dx^1\wedge dx^2\wedge
dx^3$ and $d^3x_{\mu}=d^4x(\partial_{\mu})=\frac{1}{3!}
\epsilon_{\mu\nu\rho\sigma} dx^\nu\wedge dx^\rho\wedge dx^\sigma$.  On $\Sigma$,
defined by (\ref{constr1}) and (\ref{constr2}),
\begin{equation}
\theta= - H_{0}(x^\mu,\phi^A,p_{A}^\mu)
\ d^4x + p_{A}^\mu\ d\phi^{A}\wedge d^3x_{\mu}, 
\end{equation}
and $\omega$ is the five-form
\begin{equation}
\omega= -d H_{0}(x^\mu,\phi^A,p_{A}^\mu)\wedge d^4x + d p_{A}^\mu \wedge
d\phi^A\wedge d^3x_{\mu}.
\end{equation}
An orbit of $\omega$ is a four-dimensional surface $m$ immersed in $\Sigma$,
such that at each of its points a quadruplet $X=(X_{1},X_{2},X_{3},X_{4})$ of
independent tangents to the surface satisfies
\begin{equation}
\omega(X)= 0.
\end{equation}
The projection of an orbits on $\cal C$ gives a solution of the field equations.

More in detail, let $(\partial_{\mu},\partial_{A},\partial^A_{\mu})$ be the
basis in the tangent space of $\Sigma$ determined by the coordinates
$(x^\mu,\phi^A,p_{A}^\mu)$.  Parametrize the surface with arbitrary parameters
$\tau^\rho$, so that the surface is given by
$[x^\mu(\tau^\rho),\phi^A(\tau^\rho),p^\mu_{A}(\tau^\rho)]$.  Let
$\partial_{\rho}={\partial}/{\partial\tau^\rho}$.  Then let
\begin{equation}
X_{\rho}=\partial_{\rho}x^\mu(\tau^\rho)\ \partial_{\mu}+
\partial_{\rho}\phi^A(\tau^\rho)\ \partial_{A} +
\partial_{\rho}p_{A}^\mu(\tau^\rho)\ \partial^A_\mu.
\end{equation}
Then $X=X_{0}\otimes X_{1}\otimes X_{2}\otimes X_{3}$ is a rank four tensor on
$\Sigma$.  If $\omega(X)=0$, then $\phi^A(x^\mu)$ determined by
$\phi^A(x^\mu(\tau^\rho))=\phi^A(\tau^\rho)$ is a solution of the equations of
motion.

The formalism as well as its interpretation can be immediately generalized to
the case in which the coordinates of $\cal C$ do not split into $x^\mu$ and
$\phi^A$ and the relativistic hamiltonian does not have the particular form
(\ref{constr1}-\ref{constr2}) \cite{II}.

\subsubsection*{Example: scalar field}

As an example, consider a scalar field $\phi(x^\mu)$ on Minkowski space,
satisfying the field equations
\begin{equation}
       \partial_{\mu}\partial^\mu
       \phi(x^\mu)+m^2\phi(x^\mu)+V'(\phi(x^\mu))=0.
       \label{eom}
\end{equation}
Here the Minkowski metric has signature $[+,-,-,-]$ and 
$V'(\phi)=dV(\phi)/d\phi$.  
The field is a function $\phi:M\to T$, where here $T=I\!\!R$.  The extended
configuration space of this theory is the five dimensional space with
coordinates $(x^\mu,\phi)$.  The space $\Omega$ has coordinates
$(x^\mu,\phi,\pi,p^\mu)$ (equation (\ref{constr1}) is trivially satisfied) and
carries the canonical four-form
\begin{equation}
\theta_{\Omega}= \pi\ d^4x + p^\mu\ d\phi\wedge d^3x_{\mu};
\end{equation}
The dynamics is defined on this space by the DeDonder relativistic hamiltonian
\cite{DD}
\begin{equation}
H_{0}= \frac{1}{2}\left(p^\mu p_{\mu}+m^2\phi^2 +2V(\phi)\right). 
\label{DD}
\end{equation}
The form $\omega$ is thus the five-form
\begin{equation}
\omega= -\left(p^\mu dp_{\mu}+m^2\phi d\phi +V'(\phi)d\phi\right)\wedge d^4x + d
p^\mu \wedge d\phi\wedge d^3x_{\mu}.
\end{equation}
A straightforward calculation shows that $\omega(X)=0$ gives 
\begin{eqnarray}
\partial_{\mu} \phi(x^\mu) &=& p_{\mu}(x^\mu),  \label{hfe1}\\
\partial_{\mu} p^{\mu}(x^\mu)&=& - m^2\phi(x^\mu) - V'(\phi(x^\mu)).
\label{hfe2}
\end{eqnarray}
and therefore precisely the field equations (\ref{eom}).  Notice that the
formalism is manifestly Lorentz covariant, and that no equal time initial data
surface has to be chosen.  

\subsection{The space $\cal G$ and the physical symplectic structure}

The phase space $\Gamma$ is defined as the space of the orbits, as in the finite
dimensional case.  However, notice that now there is no natural projection map
$\pi$ from $\Sigma$ to $\Gamma$, because a point in $\Sigma$ may belong to many
different orbits.  It follows that we cannot define a symplectic two-form on the
phase space $\Gamma$ by simply requiring that its pull back with $\pi$ is
$\omega$.  As we shall see now, however, the problem can be circumvented.

The key step is to identify the space $\cal G$.  Recall that in the finite
dimensional case $\cal G$ was the Cartesian product of the extended
configuration space with itself.  The same cannot be true in the field
theoretical context, because the proper characterization of $\cal G$ is as the
space of the boundary configuration data that can specify a solution.  In field
theory, we obviously need an infinite number of data to characterize a solution,
therefore $\cal G$ must be infinite dimensional.  The key observation is that in
the finite dimensional case $\cal G$ is the space of the possible
\emph{boundaries} of a portion of a motion in $\cal C$.  In the field
theoretical context, a portion of a motion is a 4d surface in $\cal C$ with
boundaries.  Its boundary is a three-dimensional surface $\gamma$.  The surface
$\gamma$ bounds a four-dimensional surface $\tilde m$, and therefore has no
boundaries itself.

Thus, we take $\cal G$ to be a space of oriented three-dimensional surfaces
$\gamma$ without boundaries in $\cal C$.  The 3d surface $\gamma$ does not need
to be connected.  In fact, it is sometimes convenient to think at $\gamma$ as
having two connected components: the initial component and the final component.

Let us coordinatize $\gamma$ with coordinates $\vec\tau =
(\tau^1,\tau^2,\tau^3)$.  Then $\gamma$ is given as
$\gamma=[x^\mu(\vec\tau),\phi^A(\vec\tau)]$.  Notice that $x^\mu(\vec\tau)$
defines a 3d surface without boundaries in Minkowski space, which we call
$\gamma_{M}$, while $\phi^A(\vec\tau)$ determines the value of the field on this
surface.  The surface in Minkowski space $\gamma_{M}$ is the boundary of a
connected region $V_{M}$ of $M$.  A solution of the equation of motion is
determined by the value of the field on the boundary.  (This is the generic
situation, since if two solutions agree on a closed 3d surface, generically they
agree in the interior.)  Thus, $\gamma$ determines a solution $\tilde m$ of the
equations of motion in the interior $V_{M}$.  Furthermore, let $m$ be the lift
of $\tilde m$ to $\Sigma$.  That is, let $m$ be the portion of an orbit of
$\omega$ that projects down to $\tilde m$.  Finally, let $s_{\gamma}$ be the 3d
surface in $\Sigma$ that bounds $m$.  That is, $s_{\gamma} = [x^\mu(\vec\tau),
\phi^A(\vec\tau), p_{A}^\mu(\vec\tau)]$, where $p_{A}^\mu(\vec\tau)$ is
determined by the solution of the field equations determined by the entire
$\gamma$.

We can now define a two-form on $\cal G$ as follows
\begin{equation}
 \omega_{\cal G}[\gamma]=\int_{s_{\gamma}} \omega.
\label{omegaGdef}
\end{equation}
The form $\omega_{\cal G}$ is a two-form: it is the integral of a five-form over
a 3d surface.  More precisely, let $\delta\gamma$ be a small variation of
$\gamma$.  This variation can be seen as a vector field $\delta\gamma(\vec\tau)$
defined on $\gamma$.  This variation determines a corresponding small variation
$\delta s_\gamma$, which, in turn, is a vector field $\delta s_\gamma(\vec\tau)$
over $s_\gamma$.  Then
\begin{equation}
 \omega_{\cal G}[\gamma](\delta_{1}\gamma,\delta_{2}\gamma) =\int_{s_{\gamma}}
 \omega(\delta_{1} s_\gamma,\delta_{2} s_\gamma).
\end{equation}
Thus, the five-form $\omega$ on the finite dimensional space $\Sigma$ defines
the two-form $\omega_{\cal G}$ on the infinite dimensional space $\cal G$.

Now, consider a small local variation $\delta\gamma$ of $\gamma$.  This means
varying the surface $\gamma_{M}$ in Minkowski space, as well as varying the
value of the field over it.  Assume that this variation satisfies the field
equations: that is, the variation of the field is the correct one, for the
solution of the field equations determined by $\gamma$.  We have
\begin{equation}
 \omega_{\cal G}[\gamma](\delta\gamma) =\int_{s_{\gamma}} \omega(\delta
 s_\gamma).
 \label{omegaG}
\end{equation}
But the variation $\delta s_\gamma$ is by construction along the orbit, namely
in the null direction of $\omega$ and therefore the right hand side of this
equation vanishes.  It follows that if $\delta\gamma$ is an infinitesimal
physical motion, then
\begin{equation}
 \omega_{\cal G}(\delta\gamma) =0.
\end{equation}

In conclusion, the pair $({\cal G},\omega_{\cal G})$ contains all the relevant
information on the system.  The null directions of $\omega_{\cal G}$ determine
the variations of the three-surfaces $\gamma$ along the physical motions.  The
space ${\cal G}$ divided by these null directions, namely the space of the
orbits of these variations is the physical phase space $\Gamma$, and the
$\omega_{\cal G}$, restricted to this space, is the physical symplectic
two-form of the system.

\subsubsection*{Example: scalar field}

Let us now compute $\omega_{\cal G}$ in a slightly more explicit form for the
example of the scalar field.  From the definition (\ref{omegaGdef}),
\begin{eqnarray}
 \omega_{\cal G}[\gamma]&=&\int_{s_{\gamma}} \omega \ \ =\ \ \int_{s_{\gamma}}
 d\pi \wedge d^4x+dp^\mu\wedge d\phi\wedge d^3x_{\mu} \nonumber \\
&=& \int_{s_{\gamma}} (p^\nu dp_{\nu}+\phi d\phi +V'd\phi)\wedge
d^4x+dp^\mu\wedge d\phi\wedge d^3x_{\mu} \nonumber \\
&=& \int_{{\gamma}_{M}} \!\!\!\! d^3x_{\nu}\, \big( (p_{\mu}\!-\!\partial_{\mu}\phi)
dp^\mu\wedge dx^\nu +(\phi\!+\!V'\!+\!\partial_{\mu}p^\mu) d\phi \wedge dx^\nu
+dp^\nu\wedge d\phi \big) \nonumber \\
&=& \int_{{\gamma}_{M}} d^3x_{\nu}\ \ dp^\nu\wedge d\phi.
\end{eqnarray}
where we have used the $x^\mu$ coordinates themselves as integration variables,
and therefore the integrand fields are the functions of the $x^\mu$'s.  Notice
that since the integral is on $s_{\gamma}$, the $p^\mu$ in the integrand is the
one given by the solution of the field equation determined by the data on
$\gamma$.  Therefore it satisfies the equations of motion
(\ref{hfe1}-\ref{hfe2}), which we have used above.  Using (\ref{hfe1}) again, we
have
\begin{eqnarray}
 \omega_{\cal G}[\gamma]= \int_{\gamma_{M}} d^3x \ n_\nu\
 d(\nabla^\nu\phi)\wedge d\phi.
\end{eqnarray}
In particular, if we consider variations $\delta\gamma$ that do not move the
surface and such that the change of the field on the surface is $\delta\phi(x)$,
we have
\begin{eqnarray}
 \omega_{\cal G}[\gamma](\delta_{1}\gamma,\delta_{2}\gamma)= \int_{\gamma_{M}}
 d^3x \ n_\nu\, \big(\delta_{1}\phi \nabla^\nu\delta_{2}\phi
 -\delta_{2}\phi\nabla^\nu\delta_{1}\phi\big).
\end{eqnarray}
This formula can be directly compared with the expression of the symplectic
two-form given on the space of the solutions of the field equations in
\cite{AW}.  The expression is the same, but with a nuance in the interpretation:
$\omega_{\cal G}$ is not defined on the space of the solutions of the field
equations -- it is defined on the space of the lagrangian data $\cal G$, and the
normal derivative $n_\nu \nabla^\nu\phi$ of these data is determined by the data
themselves via the field equations.

\subsection{Hamilton-Jacobi}

Let us now construct the function $S$ on $\cal G$. We define as in the finite 
dimensional case
\begin{equation}
S[\gamma]=\int_{m_{\gamma}} \theta.
\end{equation}
Again, it is easy to see that this is in fact the value of the action
of the solution $\tilde m_{\gamma}$.  For the scalar field, for instance 
\begin{eqnarray}
S[\gamma]&=&\int_{m_{\gamma}} \theta   
=\int_{m_{\gamma}} (\pi d^4x+p^\mu d\phi\wedge d^3x_{\mu}) 
=\int_{V_{\gamma}} \left(\pi + p^\mu \partial_{\mu}\phi\right)d^4x     
 \nonumber \\ 
&=&\int_{V_{\gamma}} \left(-\frac{1}{2}p^\mu p_\mu-\frac{1}{2}
m^2\phi^2-V(\phi)+p^\mu \partial_{\mu}\phi\right) \ d^4x 
\\ \nonumber  
&=&
\int_{V_{\gamma}}
\left(\frac{1}{2}\partial_{\mu}\phi
\partial^{\mu}\phi-\frac{1}{2}m^2\phi^2-V(\phi)\right) d^4x \nonumber \\
&=&\int_{V_{\gamma}} L(\phi,\partial_{\mu}\phi)\ d^4x,
\end{eqnarray} 
where $L$ is the Lagrangian density, and we have used the equation of 
motion $p_{\mu}=\partial_{\mu}\phi$.  

We have from the definition 
\begin{eqnarray}
\frac{\delta S[\gamma]}{\delta x^\mu(\vec\tau)}=\pi(\vec\tau)\ n_{\mu}(\vec\tau)
+\epsilon_{\mu\nu\rho\sigma}\ p^\nu(\vec\tau)\ \partial_{i}\phi(\vec\tau)\ 
\partial_{j}x^\rho(\vec\tau)\ 
\partial_{k}x^\sigma(\vec\tau)\ \epsilon^{ijk}
\label{dSx}
\end{eqnarray} 
where $\pi$ depends on the full $\gamma$, and $n_\mu(\vec\tau) = \frac{1}{3!}
\epsilon_{\mu\nu\rho\sigma} \partial_{1}x^\nu(\vec\tau) \partial_{2}
x^\rho(\vec\tau) \partial_{3} x^\sigma(\vec\tau) $ is the normal to the
three-surface $\gamma_{M}$. Also
\begin{eqnarray}
\frac{\delta  S[\gamma]}{\delta  
\phi(\vec\tau)}=p^\mu(\vec\tau)n_{\mu}(\vec\tau).
\label{dSphi}
\end{eqnarray} 
The derivation of these two equations requires steps analogous to the one we used
to derive (\ref{key}). See the appendix for details. 

Now, from (\ref{constr2}) and (\ref{DD}) we have, for the scalar field
\begin{eqnarray}
\pi+\frac{1}{2}\left(p^\mu p_{\mu}+m^2\phi^2 +2V(\phi)\right)=0. 
\end{eqnarray} 
We split $p_{\mu}$ in its normal ($p=p^\mu n_{\mu}$) and tangential
($p^i$) components (so that $p^\mu=p^i \partial_{i}x^\mu+pn^\mu$) and 
from (\ref{dSx}) we have
\begin{eqnarray}
n^{\mu}(\vec\tau) \frac{\delta S[\gamma]}{\delta x^\mu(\vec\tau)}=
\pi(\vec\tau)\ - p^i(\vec\tau)\ \partial_{i}\phi(\vec\tau). 
\end{eqnarray} 
Using this, (\ref{dSphi}), and the field equations  (\ref{dSx}), we obtain 
\begin{eqnarray}
\frac{\delta S[\gamma]}{\delta x^\mu(\vec\tau)}n_{\mu}(\vec\tau)+
\frac{1}{2}\left[\left( \frac{\delta S[\gamma]}{\delta \phi(\vec\tau)} 
\right)
^{\!\!2} \!
+\! \partial_{j}\phi(\vec\tau)\partial^j\phi(\vec\tau) +m^2 \phi^2(\vec\tau)
+2V(\phi(\vec\tau))\right]=0.
\label{HJf}
\end{eqnarray} 
This is the Hamilton-Jacobi equation of the theory. Notice that
the function $S[\gamma]=S[x^\mu(\vec\tau),\phi(\vec\tau)]$ is a
function of the surface, not the way the surface is parametrized. 
Therefore it is invariant under a change of parametrization.  It
follows that
\begin{eqnarray}
\frac{\delta  S[\gamma]}{\delta  
x^\mu(\vec\tau)}\partial_{j}x^\mu(\vec\tau)+
\frac{\delta  S[\gamma]}{\delta  
\phi(\vec\tau)}\partial_{j}\phi(\vec\tau)=0.
\label{HJf2}
\end{eqnarray} 
(This equation can be obtained also from the tangential component of
(\ref{dSx}).)  The two equations (\ref{HJf}) and (\ref{HJf2}) govern the
Hamilton-Jacobi function $S[\gamma]$.

The connection with the non-relativistic field theoretical
Hamilton-Jacobi formalism is the following.  We can restrict the formalism to a
preferred choice of parameters $\vec\tau$.  Choosing $\tau^j=x^j$, we obtain $S$
in the form $S[t(\vec x),\phi(\vec x)]$ and the Hamilton-Jacobi equation
(\ref{HJf}) becomes
\begin{eqnarray}
\frac{\delta  S}{\delta  t(\vec x)}+
\frac{1}{2}\left[\left( \frac{\delta  S[\gamma]}{\delta  
\phi(\vec x)}
\right)^2 + \partial_{j}\phi\partial^j\phi +m^2 \phi^2 
+2V(\phi)\right]=0.
\end{eqnarray} 
Further restricting the surfaces to the ones 
of constant $t$ gives the functional $S[t,\phi(\vec x)]$, 
satisfying the Hamilton-Jacobi equation 
\begin{eqnarray}
\frac{\partial  S}{\partial  t}+\frac{1}{2}\int d^3\vec x \left[\left(
\frac{\delta  S}{\delta  \phi(\vec x)} \right)^2 +
|\vec\nabla\phi|^2+m^2 \phi^2 +2V(\phi)\right]=0, 
\end{eqnarray} 
which is the usual non-relativistic Hamilton-Jacobi equation
\begin{eqnarray}
\frac{\partial S}{\partial t}+{\cal H}\left(\phi,\vec\nabla\phi,
\frac{\delta  S[\gamma]}{\delta  \phi(\vec x)}\right)=0, 
\end{eqnarray} 
where ${\cal H}(\phi,\vec\nabla\phi,\partial_{t}\phi)$ is the non-relativistic
hamiltonian.

\subsection{Physical predictions}

As in the case of finite dimensional systems, if $S[\gamma]=S[x^\mu(\vec\tau),
\phi(\vec\tau)]$ is known explicitly, the general solution of the equation of
motion can be obtained by derivations.  For instance, let $\gamma$ be formed by
two connected components that can be viewed as a past and a future Cauchy
surfaces $\gamma_{in}$ and $\gamma_{out}$, parametrized by $\vec\tau_{in}$ and
$\vec\tau_{out}$ respectively.  Consider the equation
\begin{eqnarray}
\frac{\delta S[\gamma_{out}\cup\gamma_{in}]}{\delta \phi(\vec\tau_{in})}=
\frac{\delta S[\tilde\gamma_{out}\cup\gamma_{in}]}{\delta \phi(\vec\tau_{in})}
\end{eqnarray} 
for the variable $\tilde\gamma_{out}$, where $\gamma_{in}$ and $\gamma_{out}$
are held fix.  All the solutions $\tilde\gamma_{out}$ of this equation sit on
the same motion.  That is, this equation determines which are the
$\tilde\gamma_{out}$ that are compatible with a given state.  The situation is
completely analogous to the finite dimensional case.

However, these are not the most interesting physical predictions, because
operationally well defined observables are local in spacetime.  In order to deal
with these, fix a surface $\gamma$ that determines a motion (for instance,
formed by two parallel Cauchy surfaces), and consider a single correlation
$(x^\mu,\phi^A)$ in $\cal C$.  A well posed question is whether or not the
point $(x^\mu,\phi^A)$ sits on the motion defined by $\gamma$.  That is,
whether or not the value of the field at $x^\mu$ is $\phi^A$, on the solution of
the field equations determined by the boundary conditions $\gamma$.  To answer
this question in the Hamilton-Jacobi formalism, observe that the there exist a
surface $\gamma\cup (x^\mu,\phi^A)$ in $\cal G$ and we can consider $
S[\gamma\cup (x^\mu,\phi^A)]$.  More precisely, pick a small $\epsilon$ and let
$B^\epsilon_{x^\mu}$ be a 3d surface with radius $\epsilon$ surrounding the
point $x^\mu$ in $M$.  Let $\gamma_{(x^\mu,\phi^A)}^\epsilon$ be the 3d surface 
in $\cal C$
defined by the constant value $\phi^A$ and by $x^\mu\in B^\epsilon_{x^\mu}$. 
Then
\begin{eqnarray}
S[\gamma\cup (x^\mu,\phi^A)]=\lim_{\epsilon\to 0} S[\gamma\cup
\gamma_{(x^\mu,\phi^A)}^\epsilon].
\end{eqnarray} 
Using this definition, $(x^\mu,\phi^A)$ is on the motion determined by $\gamma$ 
iff
\begin{eqnarray}
\frac{\delta S[\gamma\cup (x^\mu,\phi^A)]}{\delta \phi(\vec\tau)}=
\frac{\delta S[\gamma]}{\delta \phi(\vec\tau)}
\label{Pc}
\end{eqnarray} 
where $\vec\tau$ parametrizes $\gamma$.  This is the field theoretical
generalization of (\ref{thirdpoint2}).  This can be generalized to an arbitrary
number of correlations $(x^\mu_{1},\phi_{1}^A), \ldots, 
(x^\mu_{n},\phi_{n}^A)$.  These are compatible with the
initial data $\gamma$ iff
\begin{eqnarray}
\frac{\delta S[\gamma\cup (x^\mu_{1},\phi_{1}^A) \cup \ldots \cup 
(x^\mu_{n},\phi_{n}^A)]}{\delta \phi(\vec\tau)}=
\frac{\delta S[\gamma]}{\delta \phi(\vec\tau)}
\end{eqnarray} 
In fact, it is clear that if the correlations 
$(x^\mu_{j},\phi_{j}^A)$ are on $m$, then the insertion 
does not change the momenta on $\gamma$.  Thus, $\gamma$
determines a state, and equation (\ref{Pc}) determines the correlations in
the extended configuration space $\cal C$ that are compatible with this state. 
Clearly equation (\ref{Pc}) is the field theoretical generalization of
(\ref{thirdpoint2}).  Alternatively, we can write, as in (\ref{thirdpoint4}),
\begin{eqnarray}
\frac{\partial S[\gamma\cup (x^\mu,\phi^A)]}{\partial \phi^A}=0. 
\end{eqnarray} 

In conclusion, there are local predictions of the theory.  Given a state, the
theory can predict whether or not individual correlations (points in $\cal C$),
or sets of correlations, can be observed.

\section{General Relativity}\label{GR}

\subsection{Covariant hamiltonian formulation}

General relativity can be formulated on the finite dimensional configuration
space $\tilde{\cal C}$ with coordinates $(x^\mu, A_{\mu}^{i})$.  (See
\cite{gr4}, \cite{II} and \cite{lee}.)  Here $i=1,2,3$ and $A_{\mu}^{i}$ is a
complex matrix.  We raise and lower the $i, j, \ldots $ indices with
$\delta_{ij}$.

Assuming immediately (\ref{constr1}), the corresponding space $\Omega$ has
coordinates $(x^\mu, A_{\mu}^{i},\pi,p^{\mu\nu}_{i})$ and carries the canonical
four-form
\begin{equation}
\theta_{\Omega} = \pi \ d^4x + p^{\mu\nu}_{i}\ dA_{\nu}^{i} \wedge d^3x_{\mu}.
\label{thetaGR}
\end{equation}
It is convenient to introduce the following notation.  We define the gauge
covariant differential on all quantities with internal indices as
\begin{equation}
Dv^{i}= dv^{i}+ \epsilon^i_{jk} A_{\mu}^{j} v^{k} dx^\nu
\end{equation}
so that, in particular,
\begin{equation}
DA_{\mu}^{i}= dA_{\mu}^{i}+\epsilon^i_{jk}A_{\nu}^{j} A_{\mu}^{k} dx^\nu.
\end{equation}
Using this notation, the canonical form (\ref{thetaGR}) reads
\begin{equation}
\theta_{\Omega} = p \ d^4x + p^{\mu\nu}_{i}\ DA_{\mu}^{i} \wedge d^3x_{\nu}.
\end{equation}
where $p=\pi - p^{\mu\nu}_{i} A_{\nu}^{j} A_{\mu}^{k}\epsilon^i_{jk}$.  We
also define
\begin{equation}
E_{\mu\nu}^{i}= \epsilon_{\mu\nu\rho\sigma}\ \delta^{ij}\ p^{\rho\sigma}_{j}
\end{equation}
and the forms $A^i=A^i_{\mu}dx^\mu, DA^{i}= dA_{\mu}^{i}\wedge dx^\mu+A_{\nu
}^{j} A_{\mu}^{k}\epsilon^i_{jk} dx^\nu\wedge dx^\nu,
E^{i}=E_{\mu\nu}^{i}dx^\mu\wedge dx^\nu$, and so on, on $\Omega$.

General relativity is defined by the hamiltonian system
\begin{eqnarray}
p&=&0,\\
p_i^{\mu\nu}+p_i^{\nu\mu}&=&0,\\
\bar E^i\wedge E^j&=&0 \label{Plebanski1} \\
(\delta_{ik}\delta_{jl}-\frac{1}{3}\delta_{ij}\delta_{kl}) E^i\wedge E^j&=&0.
\label{Plebanski2}
\end{eqnarray}

Let me now show that this indeed general relativity.  The key point is that the
constraints (\ref{Plebanski1}), (\ref{Plebanski2}) imply that there exists a
real four by four matrix $e^{I}_{\mu}$, where $I=0,1,2,3$, such that
$E_{\mu\nu}^{i}$ is the self-dual part of $e^{I}_{\mu}e^{J}_{\nu}$.  This means
the following.  Let $P^i_{IJ}$ be the selfdual projector, that is
\begin{eqnarray}
P^i_{jk}&=&\epsilon^i_{jk}, \\
P^i_{j0}=-P^i_{0j}&=&i\, \delta^i_j.
\end{eqnarray}
Then it is easy to check that (\ref{Plebanski1}) and (\ref{Plebanski2}) are
solved by
\begin{eqnarray}
E^i= P^i_{IJ}\ e^I\wedge e^J.
\end{eqnarray}
and the counting of degrees of freedom indicates that this is the sole
solution.

Therefore we can use the coordinates $(x^\mu, A_\mu^i, e^I_\mu)$ on the
constraint surface $\Sigma$ (where $A_\mu^i$ is complex and $e^I_\mu$ is real)
and the induced canonical four-form is simply
\begin{equation}
\theta = P_{IJi}\ e^I \wedge e^J \wedge DA^i.
\label{thetaGR2}
\end{equation}
Indeed, the orbits $(x^\mu, A_\mu^i(x^\mu), e^I_\mu(x^\mu))$ of $\omega=d\theta$
satisfy the Einstein equations, in the form
\begin{eqnarray}
e^I\wedge (de_J+ P_{JKi}\ A^i \wedge e^K)  &=& 0, \\
P_{IJi}\ e_{I} \wedge e_{J}\wedge F^{i} &=& 0, 
\label{EE}
\end{eqnarray}
where $F_{\mu\nu}^i$ is the curvature of $A_{\mu}^i$.  From these
equations it follows that $g_{\mu\nu}(x)\equiv\eta_{IJ} e^I_\mu(x)
e^J_\nu(x)$ is Ricci flat.  The demonstration is a straightforward
calculation.

Thus, rather remarkably, the simple and natural form (\ref{thetaGR2}), defined
on the finite dimensional space $\Sigma$ with coordinates $(x^\mu, A_\mu^i,
e^I_\mu)$, defines general relativity entirely.  

\subsection{Hamilton-Jacobi equation and the $S$ solution}

Let $\gamma$ be a three-dimensional surface in $\tilde{\cal C}$.  Thus
$\gamma = [x^\mu(\vec\tau), A_{\mu}^{i}(\vec\tau)]$, where $\vec\tau =
(\tau^1,\tau^2,\tau^3)=(\tau^a)$.  Define the functional
\begin{equation}
S[\gamma]=\int_{m_{\gamma}} \theta.
\end{equation}
as above.  That is, $m$ is the four-dimensional surface in $\Sigma$ which is
(part of) an orbit of $d\theta$, and therefore a solution of the field
equations, and such that the projection of its boundary to $\tilde{\cal C}$ is
$\gamma$.  From the definition,
\begin{equation}
\frac{\delta S[\gamma]}{\delta A_{\mu}^{i}(\vec\tau)}= P_{iJK}\
\epsilon^{\mu\nu\rho\sigma}\ e^J_{\rho}(\vec\tau) e^K_{\sigma}(\vec\tau)
n_{\nu}(\vec\tau).
\label{dSee1}
\end{equation}
Since from this equation we have immediately
\begin{equation}
n_{\mu}(\vec\tau)\frac{\delta S[\gamma]}{\delta A_{\mu}^{i}(\vec\tau)}= 0,
\end{equation}
it follows that the dependence of $S[\gamma]$ on $A_{\mu}^{i}(\vec\tau)$ is only
through the restriction of $A^{i}(\vec\tau)$ to the three-surface $\gamma_{M}$. 
That is, only through the components
\begin{equation}
    A_{a}^i(\vec\tau)=\partial_{a}X^\mu(\vec\tau)A_{\mu}^i(\vec\tau).
\end{equation}
Thus $S=S[x^\mu(\vec\tau),A_{a}^i(\vec\tau)]$ and
\begin{equation}
 \frac{\delta S[\gamma]}{\delta A_{a}^{i}(\vec\tau)}= P_{iJK}\ \epsilon^{a\nu
 bc}\ 
\partial_{b}X^\rho(\vec\tau)\partial_{c}X^\sigma(\vec\tau)e^J_{\rho}(\vec\tau)
 e^K_{\sigma}(\vec\tau) n_{\nu}(\vec\tau)\equiv i E^a_{i}(\vec\tau).
\label{dSee}
\end{equation}

The projection of the field equations (\ref{EE}) on $\gamma_{M}$, written in
terms of $E^a_{i}$ read $D_{a}E^a_{i}=0$, $F_{ab}^iE^{ai}=0$ and
$F_{ab}^iE^{ai}E^{bk}\epsilon_{ijk}=0$, where $D_{a}$ and $F_{ab}^i$ are the
covariant derivative and the curvature of $A_{a}^i$.  Using (\ref{dSee}) these
give the three equations
\begin{eqnarray}
D_{a}\frac{\delta S[\gamma]}{\delta A_{a}^{i}(\vec\tau)}&=&0,  
\label{gauge}\\ 
\frac{\delta S[\gamma]}{\delta A_{a}^{i}(\vec\tau)} F_{ab}^i&=&0, 
\label{diff} \\ 
\epsilon_{ijk}F_{ab}^{i}(\vec\tau)\frac{\delta S[\gamma]}{\delta
A_{a}^{j}(\vec\tau)} \frac{\delta S[\gamma]}{\delta A_{b}^{k}(\vec\tau)}&=& 0.
\label{HJA}
\end{eqnarray}
These equations have a well known interpretation.  In fact, the first could have
been obtained by simply observing that $S[\gamma]$ is invariant under local
$SU(2)$ gauge transformations on the three-surface.  Under one such
transformation generated by a function $f^i(\vec\tau)$ the variation of the
connection is $\delta_{f}A_{a}^i=D_{a}f^i$.  Therefore $S$ satisfies
\begin{eqnarray}
0&=& \delta_{f}S=\int d^3\vec\tau\ \delta_{f}A_{a}^i(\vec\tau) \ \frac{\delta
S[\gamma]}{\delta A_{a}^{i}(\vec\tau)}=\int d^3\vec\tau\ D_{a}f^i(\vec\tau) \
\frac{\delta S[\gamma]}{\delta A_{a}^{i}(\vec\tau)}
\\ \nonumber
&=&- \int d^3\vec\tau\
f^i(\vec\tau) \ D_{a}\frac{\delta S[\gamma]}{\delta A_{a}^{i}(\vec\tau)}.
\label{variation}
\end{eqnarray}
This this gives (\ref{gauge}).  Next, the action is invariant under a change of
coordinates on the three surface $\gamma_{M}$.  Under one such transformations
generated by a function $f^a(\vec\tau)$ the variation of the connection is
$\delta_{f}A_{a}^i=f^b\partial_{b}A_{a}^i +A_{b}^i \partial_{a}f^b$. 
Integrating by parts as in (\ref{variation}) this gives
\begin{equation}
\partial_{b}A_{a}^i\frac{\delta S[\gamma]}{\delta A_{a}^{i}(\vec\tau)}
+(\partial_{b}A_{a}^i)\frac{\delta S[\gamma]}{\delta A_{a}^{i}(\vec\tau)} =0,
\end{equation}
which, combined with (\ref{gauge}) gives (\ref{diff}).  Thus, (\ref{gauge})
and (\ref{diff}) are simply the requirement that $S[\gamma]$ is invariant
under internal gauge and changes of coordinates on the three-surface.  The three
equations (\ref{gauge}),(\ref{diff}) and (\ref{HJA}) govern the dependence of
$S$ on $A_{a}^i(\vec\tau)$.

On the other hand, it is easy to see that $S$ is independent from
$x^\mu(\vec\tau)$.  A change of coordinates $x^\mu(\vec\tau)$ tangential to the
surface cannot affect the action, which is independent from the coordinates
used.  More formally, the invariance under change of parameters $\vec\tau$
implies
\begin{eqnarray}
\frac{\delta S[\gamma]}{\delta x^\mu(\vec\tau)}\partial_{j}x^\mu(\vec\tau)=
\frac{\delta S[\gamma]}{\delta A_{a}^{i}(\vec\tau)} \delta_{j}A_a^i(\vec\tau),
\end{eqnarray}
and we have already seen that the right hand side vanishes.  The variation of
$S$ under a change of $x^\mu(\vec\tau)$ normal to the surface is governed by the
Hamilton-Jacobi equation proper, Equation (\ref{HJf}). In the present case, 
following the same steps as for the scalar field, we obtain   
\begin{eqnarray}
\frac{\delta S[\gamma]}{\delta x^\mu(\vec\tau)}n_{\mu}(\vec\tau)+
\epsilon_{ijk}F_{ab}^i\frac{\delta S[\gamma]}{\delta A_{a}^{j}(\vec\tau)}
\frac{\delta S[\gamma]}{\delta A_{b}^{k}(\vec\tau)}=0.  
\label{HJAH}
\end{eqnarray}
But the second term vanishes because of (\ref{HJA}).  Therefore $S[\gamma]$ is
independent from tangential as well as normal parts of $x^\mu(\vec\tau)$: $S$
depends only on $[A_{a}^{i}(\vec\tau)]$.

We can thus drop altogether the spacetime coordinates $x^\mu$ from the extended
configuration space.  Define a smaller extended configuration space $\cal C$ as
the 9d complex space of the variables $A_{a}^{i}$.  Geometrically, this can be
viewed as the space of the linear mappings $A:D\to sl(2,C)$, where $D=R^3$ is a
``space of directions" and we have chosen the complex selfdual basis in the
$sl(2,C)$ algebra.  We then identify the space $\cal G$ as a space of
parametrized three-dimensional surfaces $[A_{a}^{i}(\vec\tau)]$ without
boundaries in $\cal C$, modulo reparametrizations -- where, however, two
parametrized surfaces are considered equivalent if $A_{a}{}^{i}(\vec\tau)=
\frac{\partial\tau'{}^b}{\partial\tau^a}A'_{b}{}^{i}(\vec\tau'(\vec\tau))$.  The
dynamics of the theory is entirely contained in the equations
(\ref{gauge}),(\ref{diff}) and (\ref{HJA}) for the functional
$S[A_{a}^i(\vec\tau)]$ on $\cal G$.

It is then immediate to obtain the dynamical equation of the quantum theory. 
This is obtained by replacing the functional derivative of $S$ with a functional
derivative operator, and acting over a wave function that has the same argument
of $S$, namely a wave functional $\Psi[A_{a}^i(\vec\tau)]$ on the extended
configuration space $\cal C$.  (A different point of view on the use of the
covariant formalism in quantum theory is developed in \cite{k}.)  Equations
(\ref{gauge}) and (\ref{diff}) do not change and demand that
$\Psi[A]=\Psi[A_{a}^i(\vec\tau)]$ is invariant under gauge transformations and
changes of coordinates $\vec\tau$, while (\ref{HJA}) becomes
\begin{eqnarray}
F_{ab}^{ij}(\vec\tau)\ \frac{\delta }{\delta A_{a}^{i}(\vec\tau)} \ \frac{\delta
}{\delta A_{b}^{j}(\vec\tau)}\ \Psi[A]=0.
\end{eqnarray}
This is the Ashtekar-Wheeler-DeWitt equation, which the basic equation of
canonical quantum gravity.  See for instance \cite{report}.

\subsection{Physical predictions}

What are the quantities predicted by the theory in the case of general
relativity?  Let $\gamma$ determine a motion $m$.  Notice that we cannot simply
ask whether a single point $c$ of $\cal C$ is in $m$, because of the non trivial
transformation properties of $A_{a}^i$ under change of parametrization.  More
precisely, if want to interpret a point in $\cal C$ as a constant field over the
3d surface of a small ball, as we did for the scalar field case, we run in the
difficulty that a constant connection on a three-sphere is trivial.  We thus
have to look for invariant extended objects in $\cal C$.  These can be defined 
as follows.  

Choose a closed unknotted curve $\alpha: s \mapsto (x^\mu(s),A_{\mu}^i(s))$ in
${\cal C}$.  Given a motion $m$, we can ask whether or not the set of
correlations forming the loop $\alpha$ can be realized.  More in general, let
$\Gamma$ be a graph (a set of points $p_{i}$ joined by lines $l_{ij}$) imbedded
in $\cal C$.  We can ask whether the collection of correlations forming $\Gamma$
is realizable in a given state, determined by a point $\gamma$ in $\cal G$.  The
answer is positive if 
\begin{eqnarray}
\frac{\delta S[\gamma]}{\delta A_a^i(\vec\tau)}=
\frac{\delta S[\gamma\cup\Gamma]}{\delta A_a^i(\vec\tau)}.
\end{eqnarray} 
Here $\vec\tau$ parametrizes $\gamma$ and $S[\gamma\cup\Gamma]$ is the integral
of $\theta$ over a motion $m_{\gamma\cup\Gamma}$ which has $\gamma$ and $\Gamma$
as boundaries.  This means $\Gamma$ is in $m_{\gamma\cup\Gamma}$.  More
precisely, we can thicken out the $M$ section of the graph $\Gamma$ as we did
for the scalar field.  Here however we do not obtain a ball of radius
$\epsilon$, but rather a sort of ``tubular structure".  The boundary of this
tubular structure is a three dimensional surface $\Gamma_{\epsilon}$ in $\cal G$
and we pose $S[\gamma\cup\Gamma]=\lim_{\epsilon\to
0}S[\gamma\cup\Gamma_{\epsilon}]$.

An important observation follows.  Consider a simple curve $\alpha$ in $\cal 
C$. 
Define now the ``holonomy" of $\alpha$
\begin{equation}
T_{\alpha}=Tr U_{\alpha}= Tr Pe^{\int_{\alpha} ds
\frac{dx^\mu(s)}{ds}A_{\mu}^i(s)\tau_{i}},
\label{Talpha}
\end{equation} 
where $\tau_{i}$ is a basis in the $su(2)$ algebra.  Let $\alpha'$ be another
closed unknotted curve in ${\cal C}$, distinct from $\alpha$, but with the same
holonomy -- that is such that $T_{\alpha'}=T_{\alpha}$.  A moment of
reflection shows that, due to the internal gauge and diffeomorphism invariance
of $S$, we have 
\begin{eqnarray}
S[\gamma\cup\alpha]=S[\gamma\cup\alpha']. 
\end{eqnarray}
Therefore the predictions of the theory do not distinguish $\alpha$ from
$\alpha'$, as far as $T_{\alpha'}=T_{\alpha}\equiv T$.  The prediction depend
only on $T$.  More in general: to any closed cycle $\alpha = l_{i_{1}i_{2}}\cup
l_{i_{p}i_{1}}$ in $\Gamma$, let the holonomy $T_{\alpha}$ be defined as in
(\ref{Talpha}).  Denote $T_{\Gamma} = (T_{\alpha_{1}}, \ldots, T_{\alpha_{n}})$
the collection of these holonomies.  Let $[\Gamma]$ be the knot-class (the
equivalence class under diffeomorphisms) to which the restriction of $\Gamma$ to
$M$ belongs, and call $s = ([\Gamma], T_{\Gamma})$ a ``spin-network".  Then to
graphs $\Gamma$ and $\Gamma'$ cannot be distinguished if they belong to the same
spin-network $s$. That is, we have 
\begin{eqnarray}
S[\gamma\cup \Gamma]=
S[\gamma\cup \Gamma']\equiv
S[\gamma\cup s]. 
\end{eqnarray} 
Therefore what the theory predicts is, for a given state,
whether or not a spin network $s$ is realizable. A set of correlations 
determined by a spin network $s$ is realizable iff 
\begin{eqnarray}
\frac{\delta S[\gamma\cup s]}{\delta A_a^i(\vec\tau)}= \frac{\delta
S[\gamma]}{\delta A_a^i(\vec\tau)}.
\end{eqnarray} 

The detection of a given $s$ can be realized in principle as follows (see also
\cite{II}).  Imagine we set up an experience in which we parallel transport a
reference system along finite paths $l_{ij}$ in spacetime.  This can be realized
macroscopically by transporting a gyroscope and a devise keeping track of local
acceleration, or, microscopically, by paralleling transport a particle with
spin. For instance a left handed neutrino, whose parallel transport is directly
described by the self-dual connection $A_{\mu}^i$.  We can then compare the
result of the parallel transport at the points $p_{i}$, thus effectively
measuring the quantities $T_{\alpha}$ as angles and relative velocities. 
\footnote{One may object that this setting can be physically realized only if
the $l_{ij}$ are all timelike and future oriented.  However, this is not a
serious experimental limitation.  First, we have obviously
$U_{l_{ij}^{-1}}=U_{l_{ij}}^{-1}$, therefore future orientation is not a
limitation in measurability.  Second, the measurement of a spacelike $l_{ij}$
can be obtained in principle as a limit of timelike ones, as in fact we do in
practice.  That is, divide $l_{ij}$ in a sequence of $N$ (spacelike) segments. 
For each such spacelike segment, bounded by the points $p_{1}$ and $p_{2}$, pick
a point $p$ in the common past of both $p_{1}$ and $p_{2}$, and consider the two
timelike geodesics that go from $p_{1}$ to $p$ and from $p$ to $p_{2}$.  Then
replace the spacelike segment with the union of these two timelike geodesics. 
It is clear that (dividing $l_{ij}$ and picking the points $p$ appropriately)
the parallel transport along the timelike curve obtained in this way converges
to the parallel transport along the spacelike curve $l_{ij}$ for large $N$. 
That is, spacelike measurements can be seen as bookkeeping for results of
measurements obtained by timelike motion.} The gauge and diffeomorphism
invariant information provided by such a measurement is then in the topology of
the graph formed by the paths and the invariant values of these relative angles
and velocities.  This is what is contained in the spin network $s$.
In the quantum theory, we expect then a quantum state to determine the
probability amplitude for any spin network $s$ to be realizable 
\cite{loops}.

\section{Conclusions and open issues}

I think that a proper understanding of the generally covariant structure of
mechanics is necessary in order to make progress in quantum gravity.  In this
paper I have made several steps in this direction.  My focus has been on
searching a physically viable language for background independent quantum field
theory and not in mathematical completeness.  From the mathematical point of
view, the structures that I have introduced in this paper certainly can (and
need to) be refined.

A result of this paper is the derivation of the symplectic structure and the
construction of the Hamilton-Jacobi formalism, in a general covariant
hamiltonian formulation of mechanics in field theory.  The main ingredient for 
this is the introduction of the space $\cal G$.  The preferred solution
$S[\gamma]$ of the Hamilton-Jacobi equation, defined on $\cal G$, contains the
full dynamical information on the system.  In finite dimensional systems, the
preferred solution $S[\gamma]$ is also the classical limit of the quantum
propagator, which contains the full dynamical information on the quantum system
in a form which makes sense in a generally covariant context, and has a direct
operational interpretation \cite{reisenberger}.  In the field theoretical
context, operationally realistic measurements are local.  Outcome of these can
be derived from $S[\gamma]$ as well.  I think that the precise relation between
$S[\gamma]$ and the Wightman amplitudes in quantum field theory deserves to be
explored.

As far as the gravitational field is concerned, the covariant hamiltonian
formulation described here is remarkably simple.  In fact, general relativity is
entirely defined on the finite dimensional space with coordinates $(x^\mu,
A_\mu^i, e^I_\mu)$, by the simple and natural form (\ref{thetaGR2}) (see also 
\cite{gr4}).  This formulation leads directly to the basic equation of quantum
gravity, and to a classical notion of spin network, which may be of help in
clarifying the physical interpretation of the quantum spin networks.

Finally, I think that there should be a proper definition of the extended
configuration space $\cal C$ in which spacetime coordinates $x^\mu$ play no role
at all, and with a clean operational interpretation.  Here and in \cite{II} I
have made some steps in this direction, but I think the matter could be further
clarified.  Ideally, I think we should get to a clean operational definition of
the partial observables of the gravitational field, and of the transition
amplitudes between correlations of these.  These are the amplitudes that a
properly covariant and background independent quantum theory of gravity should
allow us to compute.

\centerline{------}

I thank Igor Kanatchikov for useful comments, corrections on the first draft of
this work, and help with the literature on the subject.

\vskip1cm

\section*{Appendix}

Here we derive equations (\ref{key}), (\ref{dSx}) and (\ref{dSphi}).  The right
hand side of (\ref{key}) is given by the variation of the boundary in the
integral (\ref{Sgamma}).  However, this boundary variation is not the only
variation to be considered, because if $q^a$ changes, the entire curve
$m_{\gamma}$ may change, becoming a curve $m_{\gamma+\delta \gamma}$.  Thus
\begin{eqnarray}
\frac{\partial S(q^a,q^a_{0})}{\partial q^a}\delta q^a =
\int_{m_{\gamma+\delta \gamma}} \theta -
\int_{m_{\gamma}}
\theta.
\label{a}
\end{eqnarray}
Consider now the closed line integral of $\theta$ along the path
$\alpha$ formed as $\alpha=(m_{\gamma},\delta s,(m_{\gamma+\delta \gamma})^{-1}, \delta 
s_{0}^{-1})$  
\begin{eqnarray}
\oint_{\alpha}\theta=\int_{m_{\gamma}} \theta+
\int_{\delta s} \theta-\int_{m_{\gamma+\delta \gamma}} \theta - \int_{\delta 
s_{0}} \theta.
\label{b}
\end{eqnarray}
The path $\alpha$ path bounds a surface (a strip) $\sigma$ which is everywhere
tangent to the orbits of $\omega$.  Therefore the restriction of
$\omega=d\theta$ to $\sigma$ vanishes.  Therefore
\begin{eqnarray}
0=\int_{\sigma}d\theta=\int_{\alpha}\theta=0.
\label{c}
\end{eqnarray}
From (\ref{a}), (\ref{b}) and (\ref{c}), we have 
\begin{eqnarray}
\frac{\partial S(q^a,q^a_{0})}{\partial q^a}\delta q^a =\int_{\delta s} \theta-
\int_{\delta s_{0}} \theta= p_{a}\delta q^a - p_{0a}\delta q_{0}^a.
\end{eqnarray}
And since we are varying $q^a$ at fixed $q^a_{0}$, we have (\ref{key}). 

Let us now come to equation (\ref{dSx}) and (\ref{dSphi}).  Consider a surface
$\gamma$ and a small variation $\delta\gamma=(\delta x^\mu(\vec\tau),\delta
\phi(\vec\tau))$.  Consider the five dimensional strip $\sigma$ in $\Sigma$
bounded by $m_{\gamma}$,$m_{\gamma+\delta\gamma}$, and $\delta s$, where $s$ is
the boundary of $m$ and $\delta s$ its variation.  The five-form $\omega$
vanishes when restricted to $\sigma$ because $\sigma$ is tangent to the orbits
of $\omega$.  Therefore
\begin{eqnarray}
0=\int_{\sigma}\omega=\int_{\partial\sigma}\theta=
\int_{m_{\gamma+\delta\gamma}}\theta-\int_{m_{\gamma}}
\theta-\int_{\delta s}\theta. 
\end{eqnarray} 
By linearity, 
\begin{eqnarray}
\int_{m_{\gamma+\delta\gamma}}\theta-
\int_{m_{\gamma}}\theta=
\int d^3\vec \tau  
\left(\frac{\delta  S[\gamma]}{\delta  
x^\mu(\vec\tau)} \delta x^\mu(\vec\tau)
+\frac{\delta  S[\gamma]}{\delta  
\phi(\vec\tau)} \delta \phi(\vec\tau)\right).
\end{eqnarray} 
From the last two equations, (\ref{dSx}) and (\ref{dSphi}) follow with a short 
calculation.

\end{document}